\documentclass[12pt]{article}
\usepackage{graphicx}

\usepackage{slashed}
\newcommand{\be}{\begin{eqnarray}}
\newcommand{\ee}{\end{eqnarray}}
\usepackage[colorlinks]{hyperref}
\hypersetup{
    colorlinks=true,
    citecolor=blue,
    filecolor=magenta,      
    urlcolor=cyan,
}

\usepackage{amsmath,amssymb}
\usepackage{mathtools}
\usepackage{cite}
\usepackage{float}
\usepackage{color}
\usepackage{booktabs}
\usepackage{subcaption}
\usepackage{amsfonts}
\usepackage[utf8]{inputenc}
\newcommand{\mathsym}[1]{{}}
\newcommand{\unicode}[1]{{}}

\begin{document}

\allowdisplaybreaks
\begin{center}

{\Large \boldmath \bf Study of anomalous gauge-Higgs couplings using $Z$ boson polarization at LHC} 

\vskip .4cm

{\large
Kumar Rao$^a$, Saurabh D. Rindani$^b$ and Priyanka Sarmah$^a$}\\
\vskip .2cm
{$^a$\it Physics Department, Indian Institute of Technology Bombay, \\Powai,
Mumbai 400076, India}\\
\vskip .2cm
{$^b$\it Theoretical Physics Division, Physical Research Laboratory,
\\Navrangpura, Ahmedabad 380009, India} \\

\vskip 1cm
{\bf Abstract}
\end{center}
%\vskip .3cm
%\begin{abstract}
We estimate model independent bounds that could be obtained on the anomalous $ZZH$ vertex using polarization parameters of the $Z$ boson produced in the Higgstrahlung process at the LHC. We calculate the eight independent polarization parameters from the spin density matrix elements of the $Z$, which can probe underlying new physics contributions to $ZH$ production. By using the approach that connects these polarization observables to the coefficients in the angular distribution of the decay products of the $Z$, we estimate the limits on the anomalous $ZZH$ coupling that can be obtained at the 14 TeV LHC.

\section{Introduction}\label{sec1}
%With the discovery of Higgs at the LHC, the Standard Model has become the most succesful theory, endowing  explanations for most of the natural phenomena. Although experiments confirm its predictions time and again, it is very important to have precise measurements of all the properties of Higgs as well as other SM particles. Of these, the couplings of the Higgs to electroweak gauge bosons ($V=W^\pm,Z,\gamma$), $VVH$,  has a  particular importance whose form is fixed by the $SU(2)_L \times U(1)_Y$ gauge structure of the SM and are essential test of the Electroweak Symmetry Breaking mechanism, which so far has no direct experimental evidence.  Such measurements would require to go beyond simple observables like total cross section, differential rate measurements etc.The particle spectrum of the Standard Model is complete with the discovery of the Higgs boson ($H$) with mass around 125 GeV at the Large Hadron Collider (LHC). However the gauge boson sector of the model still needs  some serious attention as the exact mechanism of the   electroweak symmetry breaking is yet to be revealed. 

In the absence of evidence so far of any definitive beyond the Standard Model (SM) physics at the Large Hadron Collider (LHC), it becomes important to probe with high precision the properties of the 125 GeV Higgs ($H$) at the planned high luminosity phase of the LHC (HL-LHC). 
This requires precise measurements of the couplings of the Higgs to electroweak gauge bosons ($V=W^\pm,Z,\gamma$), its Yukawa couplings to the fermions as well as its self-couplings.  Of these, the $VVH$ couplings, whose form is fixed by the $SU(2)_L \times U(1)_Y$ gauge structure of the SM have a particular importance. Although the present scenario indicates that the couplings of the Higgs boson are in good agreement with the SM predictions, one would need more accurate measurements to further constrain the couplings or to see a small deviation from the SM predictions which could be a hint towards some underlying new physics. This will require one to go beyond usual observables like cross sections and differential rates which will be possible with higher statistics at the HL-LHC. 

A large amount of work has been carried out on probing the structure of the $VVH$ couplings at the LHC and at planned $e^+e^-$ colliders \cite{Hagiwara:1993sw, Hagiwara:2000tk, Biswal:2005fh, Godbole:2007cn, Biswal:2008tg, Biswal:2009ar, Rindani:2009pb, Rindani:2010pi, Anderson:2013afp, Godbole:2014cfa, Godbole:2013lna, Craig:2015wwr, Beneke:2014sba, Khanpour:2017cfq, Zagoskin:2018wdo, Li:2019evl, He:2019kgh}. These studies have probed the most general tensorial form of the $VVH$
coupling by using a variety of observables involving kinematic distributions of the $Z$ and the charged leptons from $Z$ decay. Study of Higgs-gauge coupling in the effective field theory framework at the LHC has been studied in \cite{Nakamura:2017ihk,Banerjee:2019pks} and at future $e^+e^-$ colliders in \cite{Chiu:2017yrx,Durieux:2017rsg ,Craig:2015wwr}. 

In this paper, we propose studying the $ZZH$ coupling by making use of the spin observables of the $Z$ boson.  We study the $ZZH$ coupling using the associated production of the $Z$ with the Higgs at the LHC. The formalism used connects various angular asymmetries of the decay products of the $Z$ to its eight independent polarization parameters extracted from the $Z$ production spin density matrix\cite{Boudjema:2009fz, Aguilar-Saavedra:2017zkn}. With the help of these parameters, we estimate limits on the anomalous couplings. $Z$ polarization has been studied in the context of new physics at the
LHC \cite{Aguilar-Saavedra:2017zkn, Nakamura:2017ihk} and at an $e^{-}e^{+}$ collider \cite{Rahaman:2017qql, Rahaman:2016pqj,K.Rao}. Analogously, polarization of the $W$ boson produced in
association with the Higgs at the LHC has been
studied in \cite{Rao:2018abz, Nakamura:2017ihk}.

The main significance of our work is that we use completely analytical
expressions for the matrix element for the production and decay of
polarized $Z$ at the partonic level. Hence the angular asymmetries that
we calculate involve no numerical calculations at the partonic level.
Only the integrations over parton distributions have to be done
numerically.

Our analytical approach has some overlap with that employed in
\cite{Nakamura:2017ihk}. However, we have estimated limits on anomalous
couplings that would be obtained at the HL-LHC, which has not been done
in \cite{Nakamura:2017ihk}.

We consider the process $pp\rightarrow Z HX$,
where  the vertex $Z_{\mu}(k_{1})\rightarrow Z_{\nu}(k_{2}) H$  has the 
Lorentz structure
\begin{equation}\label{vertex}
 \Gamma^{V}_{\mu \nu} =\frac{g}{\cos\theta_{W}}m_{Z} \left[ a_{Z}g_{\mu 
\nu}+
\frac{ b_{Z}}{m_{Z}^{2}}\left( k_{1 \nu}k_{2 \mu}-g_{\mu \nu}  k_{1}. 
k_{2}\right) +\frac{\tilde b_{Z}}{m_{Z}^{2}}\epsilon_{\mu \nu \alpha \beta} 
k_{1}^{\alpha} k_{2}^{\beta}\right]  
 \end{equation}
where $g$ is the $SU(2)_L$ coupling and $\theta_{W}$ is the weak mixing angle. $ a_{Z}$ and $ b_{Z}$  are invariant under  CP, while  $\tilde b_{Z}$ corresponds to CP violating term in the Lagrangian. In the SM, at tree level, the coupling $ a_{Z}=1$, whereas the other two couplings $b_{Z}$, $\tilde b_{Z}$ 
vanish. These vanishing couplings are the anomalous couplings  which could arise from loop corrections in the SM or in any 
extension of SM with some new particles or interactions. However, we are not concerned with the predictions of any specific model here and derive the helicity amplitudes for the process of our interest in a 
model-independent way using the general form of the $ZZH$ vertex in Eqn. (\ref{vertex}).
%\textcolor{blue}{Write on the recent constarints}

The current experimental bound on the $ZZH$ anomalous couplings is obtained by the CMS collaboration\cite{Sirunyan:2019nbs, Sirunyan:2019twz}. Although the current data are consistent with the SM predictions, the constraints are still weak enough to allow for beyond the SM contributions to the vertex. The $68\%$ confidence level (CL)  upper bounds on the $ZZH$  couplings, assuming them to be real, in our notation translate to $\vert {\rm Re }~b_Z \vert < 0.058$ and $\vert {\rm Re }~\tilde b_Z \vert < 0.078 $. These limits are obtained from measurements of ratios of the cross section contributions arising from the different $ZZH$  couplings. Ref\cite{Sahin:2019wew} obtains possible  bounds on the anomalous $ZZH$ coupling   at CLIC. For example, the $95\%$ CL limits obtained are $-0.118<b_Z  < 0.041$ and $-0.096<\tilde b_Z < 0.096$  at 3$\text{~TeV}$ centre of mass energy (c.m.) and 1000
$\text{fb}^{-1}$ integrated luminosity, neglecting systematic uncertainties. The possibility of
a future Large Hadron electron Collider (LHeC) to probe anomalous $ZZH$ 
couplings has been studied in \cite{Cakir:2013bxa}, where weak limits are
found, viz.,   $-0.21 < b_Z <  0.43$ and  $-0.32 < \tilde{b}_Z < 0.32$ 
for an electron beam energy of 60 GeV and mild improvement for a beam energy of 140 GeV, with proton beam energy of 7 TeV in either case.  

%\textcolor{blue}{Associated Higgs production with $V=W,Z$ and with $H$ decaying into $b \bar{b}$ and $V$ decaying to 0, 1 and 2 leptons has been observed by both ATLAS\cite{Aaboud:2018zhk} and CMS collaborations\cite{Sirunyan:2018kst} at close to $5\sigma$ CL. For the two-charged lepton channel $Z(l^{+}l^{-})H(b \bar{b})$ considered in our work, the significance reduces to 3.4 $\sigma$ (1.9 $\sigma$) at ATLAS (CMS) at 13 TeV cm energy and 79.8 fb$^{-1}$ (41.3 fb$^{-1}$) of data respectively. However it is seen in \cite{Goncalves:2018fvn} that how information on $Z$ boson polarization-that is not yet considered in the current collider analyses, improves  the reduced sensitivity.}
 
\section{$Z$ Polarization as a Probe}\label{sec2}

We consider the process $pp\rightarrow Z HX$ at the LHC, which at the partonic level proceeds via the process
\begin{equation}
 q(p_{1}) +\bar{q}(p_{2}) \rightarrow Z^{\alpha}(p) + H(k)
\end{equation}
through $s$-channel $Z$ exchange. Here $q$ stands for both up type and down type quarks of any generation, in the massless limit of the initial particles, with the $ZZH$ vertex given in Eqn.(\ref{vertex}). We first compute the helicity amplitudes for this process considering 
%We will later construct angular asymmetries using the charged muon from $Z$ decay. This process receives a contribution from the ``Higgsstrahlung'' diagram, mediated by a $s$-channel $Z$. Our results also hold for the $Z$ decaying to taus, to the extent that the mass of the taus can be neglected. We do not consider $Z$ decay to $e^+e^-$ to avoid interference from the SM vector boson fusion diagram,  though these effects are numerically small.
 the following representations for the 
transverse and longitudinal polarization vectors of the $Z$:
 \begin{equation}
 \epsilon^{\mu}(p,\pm)=\frac{1}{\sqrt{2}}(0,\mp \cos\theta, - i ,\pm \sin\theta),
  \end{equation}
   \begin{equation}
   \epsilon^{\mu}(p,0)=\frac{1}{m_{Z}}(\vert \vec{p}_{Z}\vert, E_{Z}\sin\theta,0,E_{Z}\cos\theta),
 \end{equation}
 where $E_{Z}$ and $\vec{p}_{Z}$ are the energy and momentum of the $Z$ 
respectively, with $\theta$ being the polar angle made by the $Z$ with respect to the quark momentum taken to be  along the positive $z$ axis.

The non-zero helicity amplitudes in the limit of massless initial 
states and assuming the SM value 
$a_Z=1$ are
\begin{eqnarray}
M(-,+,+)&=&\frac{g^{2}m_{Z}\sqrt{\hat{s}}(c_{V}+c_{A})}
{2\sqrt{2}\cos^{2}\theta_{W}(\hat{s}-m_{Z}^{2})}\left[  1- \frac{\sqrt{\hat{s}}}{ m_{Z}^{2}}(E_{Z}b_{Z}+ i  \tilde b_{Z}\vert \vec{p}_{Z}\vert)
\right]\\ \nonumber
&& \times  (1-\cos\theta)\\
M(-,+,-)&=&\frac{g^{2}m_{Z}\sqrt{\hat{s}}(c_{V}+c_{A})}
{2\sqrt{2}\cos^{2}\theta_{W}(\hat{s}-m_{Z}^{2})}\left[  1- \frac{\sqrt{\hat{s}}}{ m_{Z}^{2}}(E_{Z}b_{Z}- i  \tilde b_{Z}\vert \vec{p}_{Z}\vert)
\right] \\ \nonumber
&& \times (1+\cos\theta)\\
M(-,+,0)&=&\frac{g^{2}\sqrt{\hat{s}}(c_{V}+c_{A})}
{2\cos^{2}\theta_{W}(\hat{s}-m_{Z}^{2})}\left[E_{Z}-\sqrt{\hat{s}}b_{Z}\right]  \sin\theta \\
M(+,-,+)&=&\frac{-g^{2}m_{Z}\sqrt{\hat{s}}(c_{V}-c_{A})}
{2\sqrt{2}\cos^{2}\theta_{W}(\hat{s}-m_{Z}^{2})}\left[  1-
  \frac{\sqrt{\hat{s}}}{ m_{Z}^{2}}(E_{Z}b_{Z}+ i  \tilde b_{Z} \vert \vec{p}_{Z}\vert)\right] \\ \nonumber 
  && \times (1+\cos\theta)\\
M(+,-,-)&=&\frac{-g^{2}m_{Z}\sqrt{\hat{s}}(c_{V}-c_{A})}
{2\sqrt{2}\cos^{2}\theta_{W}(\hat{s}-m_{Z}^{2})}\left[  1-
  \frac{\sqrt{\hat{s}}}{ m_{Z}^{2}}(E_{Z}b_{Z}- i  \tilde b_{Z}\vert \vec{p}_{Z}\vert)\right] \\ \nonumber
  && \times (1-\cos\theta)\\
M(+,-,0)&=&\frac{g^{2}\sqrt{\hat{s}}(c_{V}-c_{A})}{
2\cos^{2}\theta_{W}(\hat{s}-m_{Z}^{2})}\left[E_{Z}-\sqrt{\hat{s}}b_{Z}\right]  \sin\theta
\end{eqnarray}
Here the first two entries in $M$ denote the signs of the helicities of 
the quark and antiquark respectively and the third entry is the $Z$ helicity. $\sqrt{\hat{s}}$ is
the partonic c.m. energy, and  $c_V$ and $c_A$ are the respective vector and axial vector couplings of the relevant quark to the $Z$.
 
The $a_Z$ dependence can be easily recovered
by multiplying the helicity amplitude expressions by $a_Z$,
and then replacing $b_Z$ and $\tilde b_Z$ by $b_Z/a_Z$ and $\tilde b_Z/a_Z$, 
respectively. 

We evaluate the elements of  the spin-density matrix for $Z$ production, which can be expressed in terms of the helicity amplitudes as follows
 \begin{equation}\label{rhodef}
 \rho(i,j)=\overline\sum_{\lambda,\lambda^{'}}M(\lambda,\lambda^{'},i)M^{\ast}(\lambda,\lambda^{'},j)
 \end{equation}
 the average being over the initial helicities $\lambda$, $\lambda^{'}$ 
of the quark and antiquark respectively and also over the initial color states. The $Z$ helicity indices $i,j$ can take values $\pm,0$ and with $i = j $ corresponding to the diagonal elements of Eqn.(\ref{rhodef}) which are the squared matrix elements for $Z$ production with definite polarization. It is  known that a complete information of the state of polarization is encoded in all the density matrix elements. So to attain maximum possible information, it is necessary to study the full density matrix description, which also includes the off diagonal elements.
The density matrix elements, for $q\bar{q}\to ZH$, derived from the helicity amplitudes  are given by
\begin{eqnarray}
\rho(\pm,\pm
)&=&\frac{g^{4}m^{2}_{Z}s}{8\cos^{4}\theta_{W}(\hat{s}-m_{Z}^{2})^2}
\left[(c_{V}+c_{A})^{2}(1\mp\cos\theta)^{2}\right. \nonumber\\
&& \hskip -0.8cm \left. +(c_{V}-c_{A})^{2}(1\pm\cos\theta)^{2}\right]
\left[1-2(\text{Re}~b_{Z}\mp\beta_{Z}\text{Im}~\tilde b_{Z})\frac{
E_{Z}\sqrt{\hat{s}}}{m^{2}_{Z}} \right. \nonumber\\
&& \hskip -0.8cm \left.+\frac{E_{Z}^{2}\hat{s}}{m^{4}_{Z}}\vert b_{Z}\vert^{2} \mp \frac{2E_{Z}P_{Z}\hat{s}}{m^{4}_{Z}}(\text{Im}~\tilde b_{Z}~{\rm Re}~b_{Z}-\text
{Im}~b_{Z}~{\rm Re}~\tilde b_{Z})\right. \nonumber\\
&& \hskip -0.8cm \left.+\frac{
P_{Z}^{2}\hat{s}}{m^{4}_{Z}}\vert \tilde b_{Z}\vert^{2}\right]\\
\rho(0,0
)&=&\frac{g^{4}E^{2}_{Z}s}{2\cos^{4}\theta_{W}(s-m_{Z}^{2})^2}\sin^{2}\theta
\, (c_V^2 + c_A^2) \left[1-2{\rm Re}~b_{Z}\frac{\sqrt{s}}{E_{Z}}\right. \nonumber\\
&& \hskip -0.8cm \left.
+\frac{\hat{s}}{E^{2}_{Z}}\vert b_{Z}\vert^{2}\right] \\
\rho(\pm,\mp
)&=&\frac{g^{4}m^{2}_{Z}s}{4\cos^{4}\theta_{W}(\hat{s}-m_{Z}^{2})^2}\sin^{2}\theta
 \,(c_V^2 + c_A^2)\nonumber \\ &&\times 
\left[1-2({\rm Re}~b_{Z}\pm  i \beta_{Z} {\rm Re}~\tilde b_{Z})\frac{
E_{Z}\sqrt{s}}{m^{2}_{Z}}+\frac{
E_{Z}^{2}\hat{s}}{m^{4}_{Z}}\vert b_{Z}\vert^{2}\pm i \frac{2E_{Z}P_{Z}\hat{s}}{m^{4}_{Z}}\right.\nonumber\\
&& \hskip -0.8cm \left.(\text{Im}~\tilde b_{Z}~\text
{Im}~b_{Z}+{\rm Re}~b_{Z}~{\rm Re}~\tilde b_{Z})-\frac{
2P_{Z}^{2}\hat{s}}{m^{4}_{Z}}\vert \tilde b_{Z}\vert^{2} \right] \\ \nonumber
\rho(\pm,0 )&=&\frac{g^{4}m_{Z}E_{Z}s}
{4\sqrt{2}\cos^{4}\theta_{W}(\hat{s}-m_{Z}^{2})^2}\sin\theta
\nonumber \\
&& \hskip -.8cm \times \left[
(c_{V}+c_{A})^{2}(1\mp\cos\theta) 
 -(c_{V}-c_{A})^{2}(1\pm\cos\theta)\right] \nonumber\\
&& \hskip -.8cm \times 
\left[1-{\rm Re}~b_{Z}\sqrt{\hat{s}}\frac{
(E^{2}_{Z}+m^{2}_{Z})}{E_{Z}m^{2}_{Z}}
 - i \sqrt{\hat{s}} \frac{E_{Z}}{m^{2}_{Z}}\left({\rm
Im}~b_{Z}~\beta^{2}_{Z}\pm \tilde
b_{Z} \beta_{Z}\right) \right. \nonumber\\
&& \hskip -0.8cm \left.
\mp \frac{\hat{s}}{m^{2}_{Z}}\vert b_{Z}\vert^{2}\pm\frac{\hat{s}P_{Z}}{m^{2}_{Z}E_{Z}} ({\rm Im}~b_{Z}+i {\rm Re}~b_{Z})({\rm Re}~\tilde b_{Z}+i \text{Im}~\tilde b_{Z})\right] 
\end{eqnarray}
where $\beta_{Z}=\vert\vec{p}_{Z}\vert /E_{Z}$ is the velocity of the $Z$ in 
the c.m frame. The analytical manipulation software FORM 
\cite{Vermaseren:2000nd} has been used to verify these expressions.
We have kept the finite $Z$ width 
in our numerical calculations later.

The full density matrix Eqn.(\ref{rhodef}) on  integrating over an appropriate kinematic range, can be parametrized in terms of the 3 components of the vector polarization $\vec P$ and 5 components of the tensor polarization $T$   of the $Z$ boson\cite{Leader:2001gr}. Defining this as $\sigma(i,j)$ we have
\begin{equation}\label{vectensorpol}
\sigma(i,j) \equiv \sigma \;\left( 
\begin{array}{ccc}
\frac{1}{3} + \frac{P_z}{2} + \frac{T_{zz}}{\sqrt{6}} &
\frac{P_x - i P_y}{2\sqrt{2}} + \frac{T_{xz}-i T_{yz}}{\sqrt{3}} &
\frac{T_{xx}-T_{yy}-2iT_{xy}}{\sqrt{6}} \\
\frac{P_x + i P_y}{2\sqrt{2}} + \frac{T_{xz}+i T_{yz}}{\sqrt{3}} &
\frac{1}{3} - \frac{2 T_{zz}}{\sqrt{6}} &
\frac{P_x - i P_y}{2\sqrt{2}} - \frac{T_{xz}-i T_{yz}}{\sqrt{3}} \\
\frac{T_{xx}-T_{yy}+2iT_{xy}}{\sqrt{6}} &
\frac{P_x + i P_y}{2\sqrt{2}} - \frac{T_{xz}+i T_{yz}}{\sqrt{3}} &
\frac{1}{3} - \frac{P_z}{2} + \frac{T_{zz}}{\sqrt{6}} 
\end{array}
\right)
\end{equation}
where  $\sigma$ is the 
production cross section, 
\begin{equation}
\sigma = \sigma(+,+) + \sigma(-,-) + \sigma(0,0).
\end{equation}
The eight independent vector and tensor polarization observables of the $Z$ can then be constructed using appropriate linear combinations of the integrated density matrix elements of Eqn.(\ref{vectensorpol}):
\begin{eqnarray}\label{pol1}
P_{x}&=&  \frac{\lbrace \sigma(+,0)+\sigma(0,+)\rbrace
 +\lbrace \sigma(0,-)+\sigma(-,0)\rbrace}{\sqrt{2}\sigma}\\ \label{pol2}
 P_{y}&=&\frac{- i  \lbrace[\sigma(0,+)-\sigma(+,0)]+[\sigma(-,0)-\sigma(0,-)]\rbrace}{\sqrt{2}\sigma}\\\label{pol3}
 P_{Z}&=&\frac{[\sigma(+,+)]-[\sigma(-,-)]}{\sigma}\\ \label{pol4}
 T_{xy}&=&\frac{- i  \sqrt{6}[\sigma(-,+)-\sigma(+,-)]}{4\sigma}\\ \label{pol5}
  T_{xz}&=&\frac{\sqrt{3}\lbrace[\sigma(+,0)+\sigma(0,+)]-[\sigma(0,-)+\sigma(-,0)]\rbrace}{4\sigma}\\ \label{pol6}
  T_{yz}&=&\frac{- i  \sqrt{3}\lbrace[\sigma(0,+)-\sigma(+,0)]-[\sigma(-,0)-\sigma(0,-)]\rbrace}{4\sigma}\\ \label{pol7}
  T_{xx}-T_{yy}&=&\frac{\sqrt{6}[\sigma(-,+)+\sigma(+,-)]}{2\sigma}\\ \label{pol8}
  T_{zz}&=&\frac{\sqrt{6}}{2}\left\lbrace\frac{[\sigma(+,+)]+[\sigma(-,-)]}{\sigma}-\frac{2}{3}\right\rbrace
\nonumber \\
 &=&\frac{\sqrt{6}}{2}\left[\frac{1}{3}-\frac{\sigma(0,0)}{\sigma}\right] \label{pol9}
 \end{eqnarray}
Of these $P_x$, $P_y$ and $P_z$ are the vector polarizations, whereas
the $T$'s are the tensor polarizations, with the constraint that the
tensor is traceless.
% In the following section we give relations of these to 
%angular asymmetries of the decay leptons of the $Z$ which would serve as measures
%of the polarization parameters.}
%\section{Lepton asymmetries}\label{sec3}
In real experiments where the $Z$ boson decays to two leptons, these polarization observables can be extracted from  kinematic
distributions of its decay products. Angular asymmetries  can be obtained by combining the relevant production-level density matrix elements with appropriate decay density matrix elements and integrating over the appropriate phase space.  For example, $P_x$ can be calculated from the asymmetry $A_x$ defined by-

\begin{equation}\label{asyx}
%\centering
A_{x}=\frac{3\alpha P_{x}}{4}\equiv\frac{\sigma(\cos\phi^{\ast}>0)-\sigma(\cos\phi^{\ast}<0)}{\sigma(\cos\phi^{\ast}>0)+\sigma(\cos\phi^{\ast}<0)}
\end{equation}
where, $\alpha$ is the $Z$ boson polarization analyzer, given in terms of its vector and axial vector couplings to charged leptons $\ell$, $c_V^\ell $ and $c_A^\ell $ respectively, as
\begin{equation}\label{polanalyser}
\alpha =-\frac{2 c_V^\ell  c_A^\ell}{{c_V^{\ell}}^2 +{c_A^{\ell}}^2}
\end{equation}
%\frac{R_\ell^2 -L_\ell ^2}{R_\ell^2 +L_\ell ^2}=
The angles $\theta^\ast$ and $\phi^\ast$ are polar and azimuthal angles of the lepton in the rest frame of the $Z$. The $Z$ rest frame is reached by  a combination of boosts and rotations from the laboratory frame. In the laboratory frame, the quark momentum defines the positive $z$ axis, and the production plane of $Z$ is
defined as the $xz$ plane. While boosting to the $Z$ rest frame, the $xz$
plane is kept unchanged. Then, the angles $\theta^\ast$ and $\phi^\ast$
are  measured with respect to the would-be momentum of the $Z$.
Similarly, expressions for other asymmetries, \textit{viz.},
$A_{y}$, $A_{z}$, $A_{xy}$, $A_{yz}$, $A_{xz}$, $A_{x^{2}-y^{2}}$, $A_{zz}$
corresponding to the 2 vector polarizations $P_y$, $P_z$  and 5 tensor polarizations $T_{ij}(i,j=x,y,z)$ can be obtained and are listed in \cite{Rahaman:2016pqj, K.Rao}. 
%But a large number 
%of events is required for full measurement of the lepton distribution. Therefore 
%a more economic way is to use the integrated angular asymmetries, which utilize 
%all relevant events. 

 It is observed that the density matrix elements $\sigma(\pm,0)$ and $\sigma(0,\pm)$ and
the asymmetries involving these elements $A_{x}$, $A_y$, $A_{xz}$, $A_{yz}$ vanish due to the symmetric
nature of the LHC, which does not allow a unique definition of a positive $z$-axis. Therefore to make them non-zero, we define the direction of the
reconstructed momentum of the  $ZH$ combination as the positive $z$-axis.

We have evaluated these asymmetries upto quadratic order in the anomalous couplings.
It is observed that out of eight polarization asymmetries only 
three, {\it viz.}, $A_{x}$, $A_{x^{2}-y^{2}}$ and $A_{zz}$ are non-zero in the 
SM, which,  along 
with the total cross section, would be proportional to the real part of the 
anomalous couplings (upto linear order) or absolute square of the couplings to satisfy the CPT theorem. This will be seen in the following 
section.
\section{Limits on the Anomalous Couplings}
\label{sec4}

Here we present numerical values for the integrated density matrix
elements, the corresponding asymmetries and the sensitivities of the
asymmetries to the various anomalous 
couplings. We consider 
c.m. energy $\sqrt{s}=14\text{~TeV}$, with integrated luminosity
$\int \mathcal{L}dt=1000$ $\text{fb}^{-1}$. In our numerical
calculations, we employ MMHT2014 parton distribution functions
\cite{mmht} with factorization scale
chosen as the square root of the partonic c.m. energy.
The integrated production density matrix elements in the units of fb are
\begin{eqnarray}
\sigma(\pm,\pm)&=& 161.95 - 1495.62~\text{Re}~b_{Z} \pm 1036.98 ~\text{Im}~\tilde{b}_{Z}+5391.21 \, \vert b_{Z}\vert^{2} \nonumber \\
&& \hskip -0.8cm + 3753.23 \, \vert\tilde b_{Z}\vert^{2}\mp 
 8811.36~ (\text{Im}~\tilde{b}_{Z} ~\text{Re}~b_{Z}-~\text{Im}~b_{Z} ~\text{Re}~\tilde{b}_{Z}) \\ \nonumber \\
\sigma(0,0)&=&341.976  -1495.62 ~\text{Re}~b_{Z}+  1637.98\, \vert b_{Z}\vert^{2} \\ \nonumber \\
\sigma(\pm,\mp)&=&  80.97 - 747.81 ~\text{Re}~b_{Z}\mp i 518.49  ~\text{Re}~\tilde{b}_{Z}+2695.6\vert b_{Z}\vert^{2}\nonumber \\
&& \hskip -0.8cm 
-  1876.61\, \vert\tilde b_{Z}\vert^{2}\pm i 4405.67 ~(\text{Im}~\tilde{b}_{Z} ~\text{Im}~b_{Z}+~\text{Re}~\tilde{b}_{Z} ~\text{Re}~b_{Z}) \\\nonumber \\
\sigma(\pm,0)&= &59.59 - 474.46 ~\text{Re}~b_{Z}-i 211.22 ~\text{Im}~b_{Z}\mp 261.88  ~(i ~\text{Re}~\tilde{b}_{Z}- \text{Im}~\tilde{b}_{Z})\nonumber \\
&& \hskip -0.8cm 
+738.07\, \vert b_{Z}\vert^{2} \pm 
558.15~ \tilde{b}_{Z}~(\text{Im}~b_{Z}+i ~\text{Re}~b_{Z}) \\\nonumber \\
\sigma(0,\pm)&=& 59.59 - 474.46 ~\text{Re}~b_{Z}+i 211.22 ~\text{Im}~b_{Z}\pm 261.88  ~(i~\text{Re}~\tilde{b}_{Z}+\text{Im}~\tilde{b}_{Z})\nonumber \\
&& \hskip -0.8cm 
+738.07\, \vert b_{Z}\vert^{2} \pm 
558.15~\tilde{b}_{Z}~(\text{Im}~b_{Z}+i ~\text{Re}~b_{Z})
\end{eqnarray}
The leptonic asymmetries corresponding to different polarizations calculated  upto linear order in the anomalous couplings  are given by
\begin{equation}
A_{x}=0.035 ~\text{Re}~b_{Z}-0.028
\end{equation}

\begin{equation}
A_{y}=-0.125~ \text{Re}~\tilde{b}_{Z}
\end{equation}

\begin{equation}
A_{z}=-0.349~ \text{Im}~\tilde{b}_{Z}
\end{equation}

\begin{equation}
A_{xy}=0.496~ \text{Re}~\tilde{b}_{Z}
\end{equation}

\begin{equation}
A_{xz}=-0.354 ~\text{Im}~\tilde{b}_{Z}
\end{equation}

\begin{equation}
A_{yz}=0.286 ~\text{Im}~b_{Z}
\end{equation}

\begin{equation}
A_{x^{2}-y^{2}}=-0.193 ~\text{Re}~b_{Z}+0.077
\end{equation}

\begin{equation}
A_{zz}=-0.683~ \text{Re}~b_{Z}-0.101
\end{equation}

It is observed that all asymmetries except $A_{x}$, $A_{x^{2}-y^{2}}$ and $A_{zz}$
vanish in the SM and the reason for this is the CP even  and T even nature of the asymmetries $A_x, \, A_{x^2-y^2}$ and $A_{zz}$, because of which they can occur at tree level in the SM. The remaining asymmetries vanish in the SM because they are either CP even and T odd  or CP odd, and hence depend on the  CP violating parameters which are absent in the SM at tree level. 
\begin{table}[H]
\centering
\begin{tabular}{|c|c|c|}
%\begin{tabular}{ |c|c|p{2cm}|p{2.2cm}|}
% \hline
   \hline
%& & \multicolumn{2}{|c|}{Limit ($\times 10^{-3}$) for } \\
% \hline
 %\hline
 Observable &Coupling &  Limit ($\times 10^{-3}$)  \\
 
\hline
%\hline
$\sigma$   & $\text{Re}~b_{Z}$  & $0.70$\\
%\hline
$A_{x}$& $\text{Re}~b_{Z}$  & $136 $\\
%\hline
$A_{y}$ & $\text{Re}~\tilde{b}_{Z}$&   $ 37.9 $\\
%\hline
$A_{z}$   & $ \text{Im}~\tilde{b}_{Z}$&   $ 13.5$\\
%\hline&$x$& $x$\\
% \hline
$A_{xy}$   & $\text{Re}~\tilde{b}_{Z}$  &   $9.53 $\\
% \hline
 $A_{yz}$   & $\text{Im}~b_{Z}$  &   $16.5 $\\
% \hline
 $A_{xz}$   & $ \text{Im}~\tilde{b}_{Z}$  &   $ 13.3 $\\
% \hline
 $A_{x^{2}-y^{2}}$   & $\text{Re}~b_{Z}$  &   $ 24.4 $\\
%\hline
 $A_{zz}$   & $\text{Re}~b_{Z}$  &   $ 6.88 $\\
\hline
\end{tabular}
\caption{$1 \sigma$ limit obtained on the anomalous couplings from cross section and  various leptonic asymmetries calculated upto linear order in the couplings 
at $\sqrt{s}=14 \text{~TeV}$.  }
\label{table:9}
\end{table}
Next we present the expressions for the total cross section and  angular asymmetries including quadratic terms of couplings at $\sqrt{s}=14 \text{~TeV}$.
\begin{equation}
\sigma=0.067294~ (7506.45~ \vert\tilde b_{Z}\vert^{2}+12420.4 ~\vert b_{Z}\vert^{2}-4486.85~ \text{Re}~b_{Z}+665.87)~\text{fb}
\end{equation}

\begin{equation}
A_{x}=\frac{0.012~\text{Re}~b_{Z} -0.019~ \vert b_{Z}\vert^{2}-0.002}{0.604~ \vert\tilde b_{Z}\vert^{2}+\vert b_{Z}\vert^{2}-0.361 ~\text{Re}~b_{Z}+0.054}
\end{equation}

\begin{equation}
A_y=\frac{0.024~ \text{Im}~\tilde{b}_{Z}~ \text{Im}~b_{Z}+\text{Re}~\tilde{b}_{Z}~(0.024  ~\text{Re}~b_{Z}-0.011)}{\vert\tilde b_{Z}\vert^{2}+1.655~ \vert b_{Z}\vert^{2}+(\text{Re}~\tilde{b}_{Z})^2-0.598~ \text{Re}~b_{Z}+0.089}
\end{equation}

\begin{equation}
A_z=\frac{\text{Im}~\tilde{b}_{Z} ~(1976.66~ \text{Re}~b_{Z}-232.627)-1976.66 ~\text{Im}~b_{Z} ~\text{Re}~\tilde{b}_{Z}}{7506.45~  \vert\tilde b_{Z}\vert^{2}+12420.4 ~\vert b_{Z}\vert^{2}-4486.85~ \text{Re}~b_{Z}+665.87}
\end{equation}

\begin{equation}
A_{xy}=\frac{\text{Re}~\tilde{b}_{Z} ~(0.044-0.374~ \text{Re}~b_{Z})-0.374~ \text{Im}~\tilde{b}_{Z}~\text{Im}~b_{Z}}{ \vert\tilde b_{Z}\vert^{2}+1.655~\vert b_{Z}\vert^{2}-0.598~ \text{Re}~b_{Z}+0.089}
\end{equation}

\begin{equation}
A_{yz}=\frac{190.164~\text{Im}~b_{Z}}{7506.45~  \vert\tilde b_{Z}\vert^{2}+12420.4 ~\vert b_{Z}\vert^{2}-4486.85~ \text{Re}~b_{Z}+665.87}
\end{equation}

\begin{equation}
A_{xz}=\frac{\text{Im}~\tilde{b}_{Z} ~(0.0404 ~\text{Re}~b_{Z}-0.019)-0.0404~\text{Im}~b_{Z} ~\text{Re}~\tilde{b}_{Z}}{0.604~ \vert\tilde b_{Z}\vert^{2}+\vert b_{Z}\vert^{2}-0.361 ~\text{Re}~b_{Z}+0.054}
\end{equation}

\begin{equation}
A_{x^{2}-y^{2}}=\frac{-0.297~ \vert\tilde b_{Z}\vert^{2}+0.019~\text{Re}~b_{Z}-0.005}{ \vert\tilde b_{Z}\vert^{2}+1.655~\vert b_{Z}\vert^{2}-0.598~ \text{Re}~b_{Z}+0.089}+0.138
\end{equation}

\begin{equation}
A_{zz}=\frac{0.074~ \vert\tilde b_{Z}\vert^{2}+0.068 ~\text{Re}~b_{Z}-0.019}{ \vert\tilde b_{Z}\vert^{2}+1.655~\vert b_{Z}\vert^{2}-0.598~ \text{Re}~b_{Z}+0.089}+0.113
\end{equation}

We obtain  the sensitivity of an observable $\mathcal{O}$
which depends on a parameter $f$ from the definition
%\begin{equation}\label{limit-sys}
%C_{\rm{limit}}=\frac{1}{\vert A-A_{SM}\vert}\sqrt{\frac{1-A^{2}_{SM}}
%{\sigma_{SM}\mathcal{L}} + \epsilon_A^2 },
%\end{equation}
\begin{equation}\label{limit}
\mathcal{S}(\mathcal{O}(f))=\frac{\vert\mathcal{O}(f)-\mathcal{O}(f=0) \vert}{\delta \mathcal{O}}
\end{equation}
where $\delta \mathcal{O}$ is the estimated error on the
observable. For an asymmetry, the estimated error takes the form 
\begin{equation}\label{limit_A}
\delta A=\frac{\sqrt{1-A^{2}_{SM}}}{\sqrt{\sigma_{SM}\mathcal{L}}}
\end{equation}
%\textcolor{red}{ 
with $\sigma_{SM}$ being the SM cross section for the process $pp \to Z^*H \to \ell\bar{\ell}H$ ($\ell= e, \mu$) at the LHC with integrated luminosity $\mathcal{L}$ and
$A_{SM}$ is the corresponding value of asymmetry in the SM. Similarly, for the cross section, the error is given by
%}
 \begin{equation}\label{limit-cs}
\delta \sigma=\sqrt{\frac{\sigma_{\rm{SM}}}{\mathcal{L}}}
\end{equation}
\newcommand{\Rebzt}{Re~$\tilde b_Z$~}
\newcommand{\Imbzt}{Im~$\tilde b_Z$~}
\newcommand{\Rebz}{Re~$b_Z$~}
\newcommand{\Imbz}{Im~$b_Z$~}
We  estimate the 1$\sigma$ limits calculated upto linear order  and list it in Table \ref{table:9}. We note that, among the four observables $\sigma$, $A_{x}$, $A_{zz}$ and $A_{x^{2}-y^{2}}$, which are sensitive to \Rebz\!\!, the total cross section provides the best limits on the coupling. However, it is not sufficient to consider 
the total cross sections as the only probe as it is sensitive to just one  coupling,
\Rebz\!\!. So to explore the couplings which do not appear in the total cross section, one will require the other angular asymmetries.
The better limit on \Imbzt comes from $A_{x}$ and $A_{xz}$, both being equally sensitive to the coupling.  For the coupling \Imbz\!\!, the best bound comes from $A_{yz}$ whereas  on \Rebzt  the best limit of $9.53\times 10^{-3} $ is achieved from the observable $A_{xy}$.

\begin{figure}[H]
\begin{subfigure}{.55\textwidth}
 \centering
  \includegraphics[width=.9\linewidth]{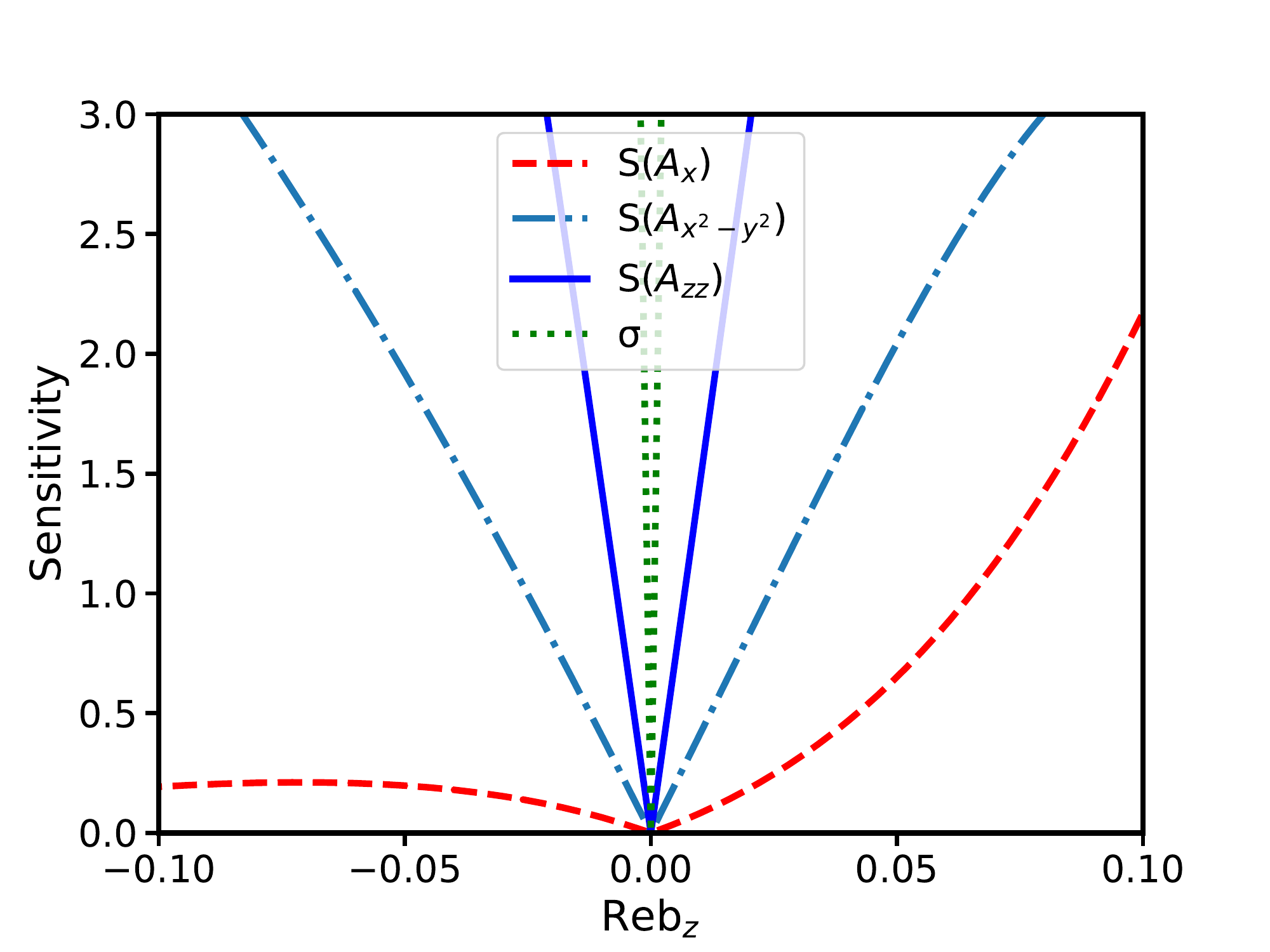}
 % \caption{}
  \label{fig: 1}
\end{subfigure}%
\begin{subfigure}{.55\textwidth}
  \centering
  \includegraphics[width=.9\linewidth]{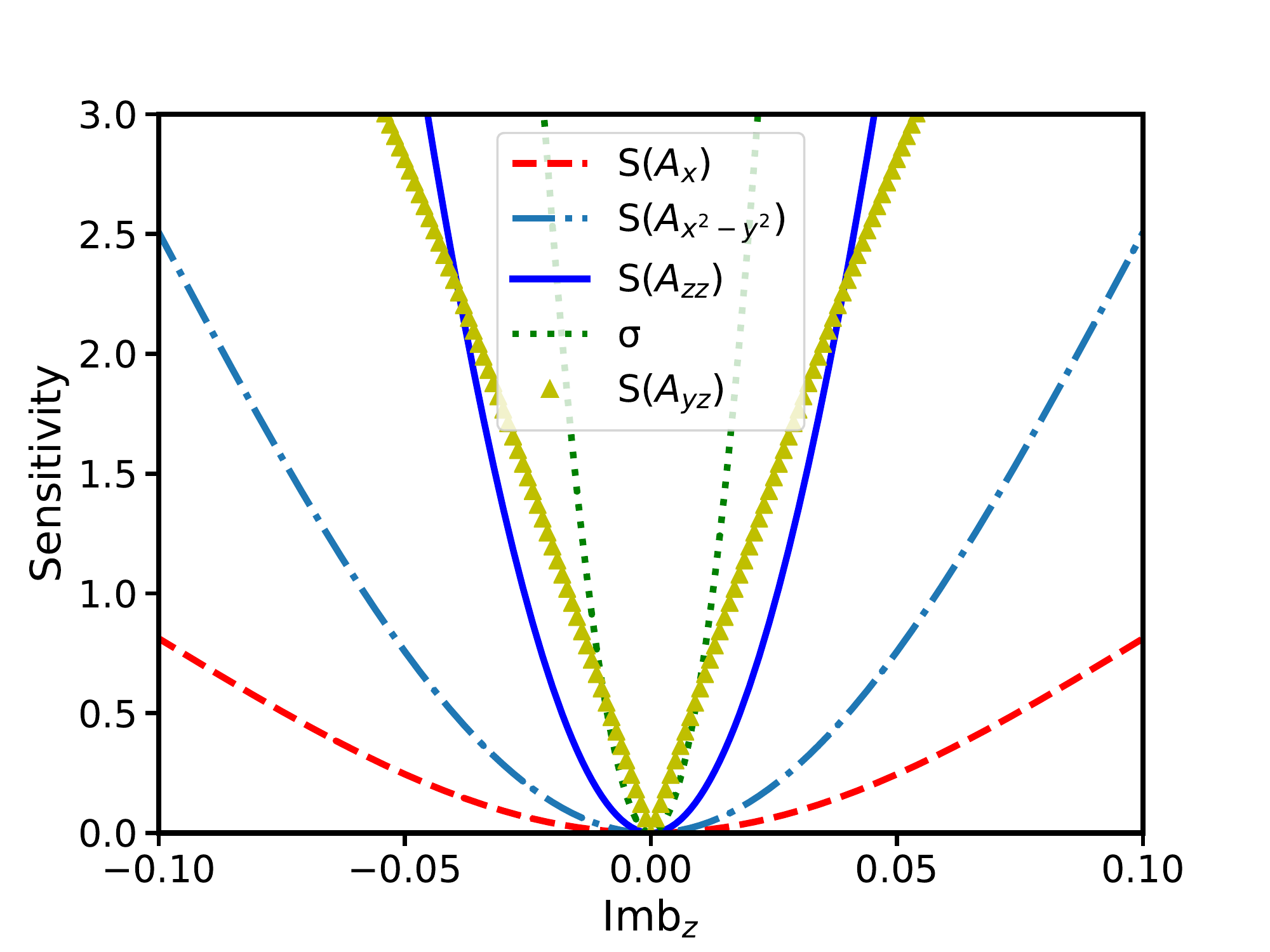}
  %\caption{}
  \label{fig: 2}
\end{subfigure}

\begin{subfigure}{.55\textwidth}
 \centering
  \includegraphics[width=.9\linewidth]{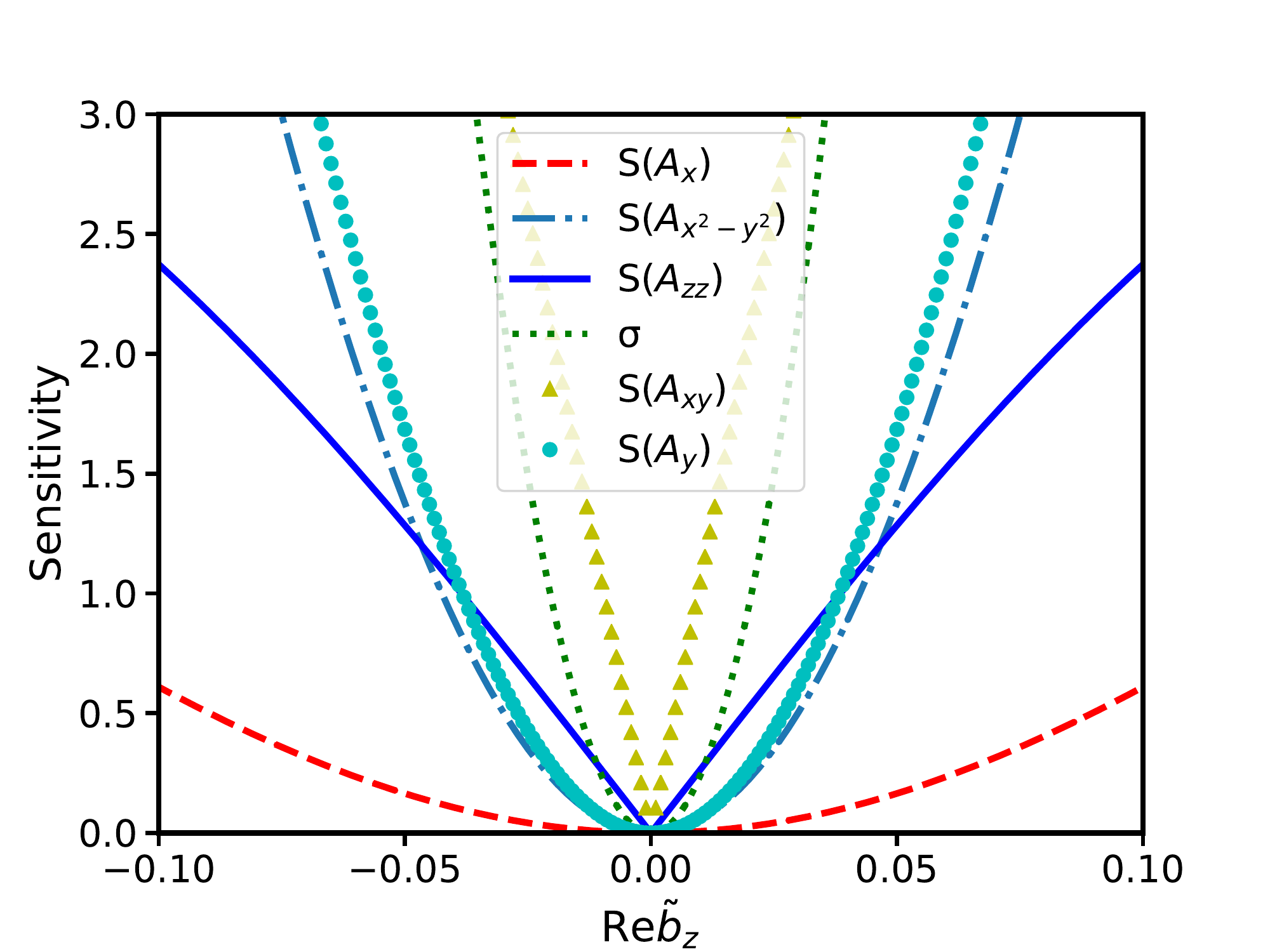}
 % \caption{}
  \label{fig: 3}
\end{subfigure}%
\begin{subfigure}{.55\textwidth}
  \centering
  \includegraphics[width=.9\linewidth]{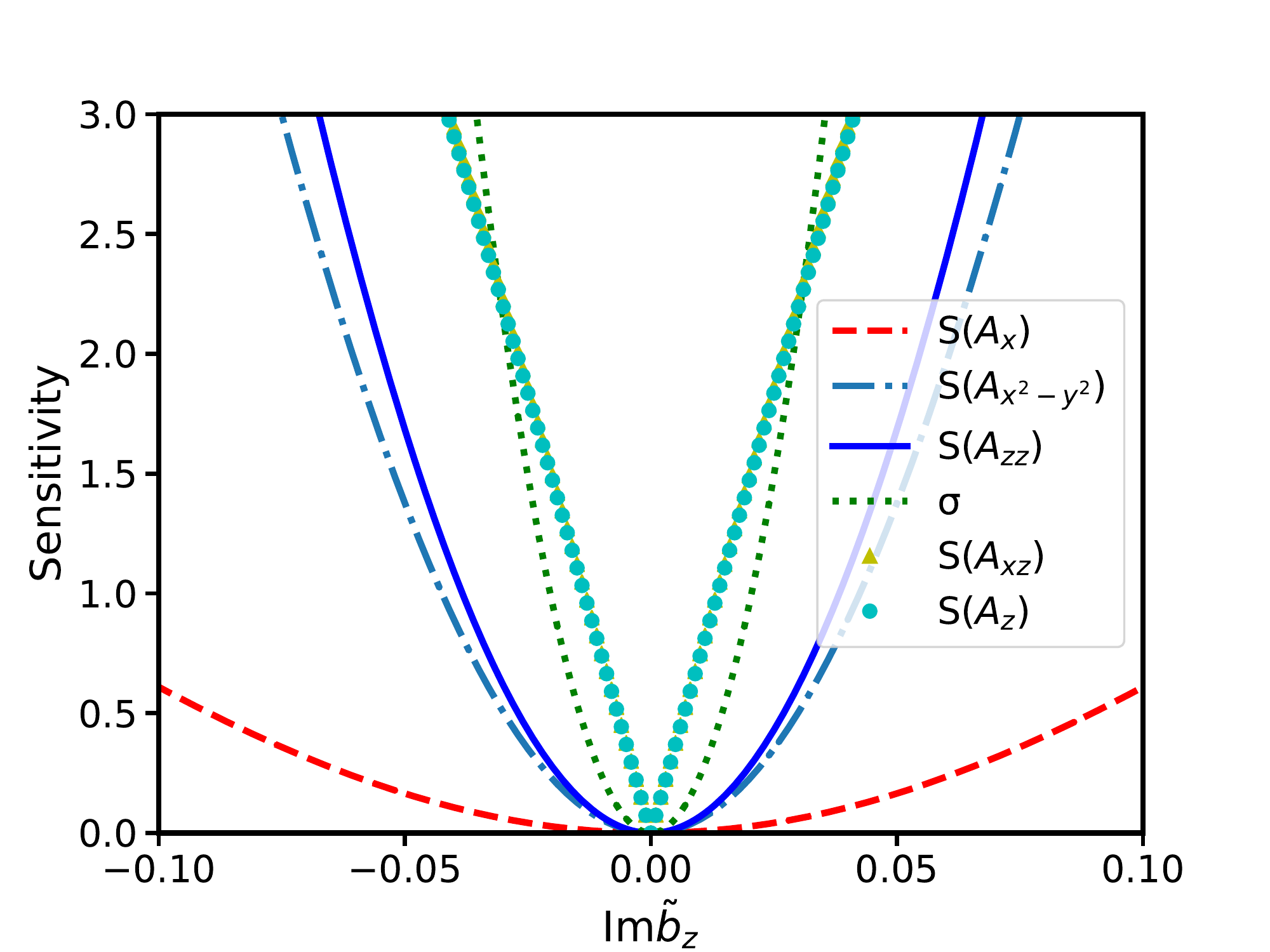}
%  \caption{}
  \label{fig: 4}
\end{subfigure}

\caption{Sensitivities of cross section and asymmetries to anomalous couplings, including quadratic order at $\sqrt{s}=14$TeV. Plots are obtained by varying one coupling at a time.}
\label{fig:1}
\end{figure}   
\begin{table}[H]
\centering
\begin{tabular}{|c|c|c|}
%\begin{tabular}{ |c|c|p{2cm}|p{2.2cm}|}
% \hline
   \hline
%& & \multicolumn{2}{|c|}{Limit ($\times 10^{-3}$) for } \\
% \hline
 %\hline
 Observable &Coupling &  Limit ($\times 10^{-3}$)  \\
 
\hline
%\hline
 $\sigma$  & $\vert\text{Re}~b_{Z}\vert$  & $0.70$\\
%\hline
$\sigma$ & $\vert\text{Im}~b_{Z}\vert$  & $15.9 $\\
%\hline
$A_{xy}$ & $\vert\text{Re}~\tilde{b}_{Z}\vert$&   $ 9.54 $\\
%\hline
$A_{xz}$, $A_{z}$   & $\vert\text{Im}~\tilde{b}_{Z}\vert$&   $  13.3$\\
\hline
\end{tabular}
\caption{The  best $1 \sigma$ limit  on couplings and 
the corresponding observables at $\sqrt{s}=14\text{~TeV}$, obtained  from Figure \ref{fig:1}.}
\label{table:2}
\end{table}

In Figure \ref{fig:1}, we plot the one parameter sensitivity $\textit{i.e}$  $\mathcal{S}=1$(or $\Delta\chi^2=1$) for the cross section and the 8 asymmetries, considered upto quadratic order in the anomalous couplings. It is observed from Figure \ref{fig:1} that the tightest limit on the coupling $\text{Re}~b_{Z}$ can be obtained 
from total 
cross section. On the coupling $\text{Im}~b_{Z}$,  both cross section
and  $A_{yz}$ place comparable limits.
%, but limits from cross section gets weaker for higher values of sensitivity. 
The observables $A_{x}$ and $A_{xz}$ are found to be almost equally sensitive to  the coupling $ \text{Im}~\tilde{b}_{Z}$. The best limit on $\text{Re}~\tilde{b}_{Z}$ can be obtained from the observable  $A_{xy}$. In Table \ref{table:2}, we list the tightest $1 \sigma$ level limit on the couplings, obtained from Figure \ref{fig:1}.
\begin{figure}[] % "[t!]" placement specifier just for this example
\begin{subfigure}{0.55\textwidth}
\includegraphics[width=\linewidth]{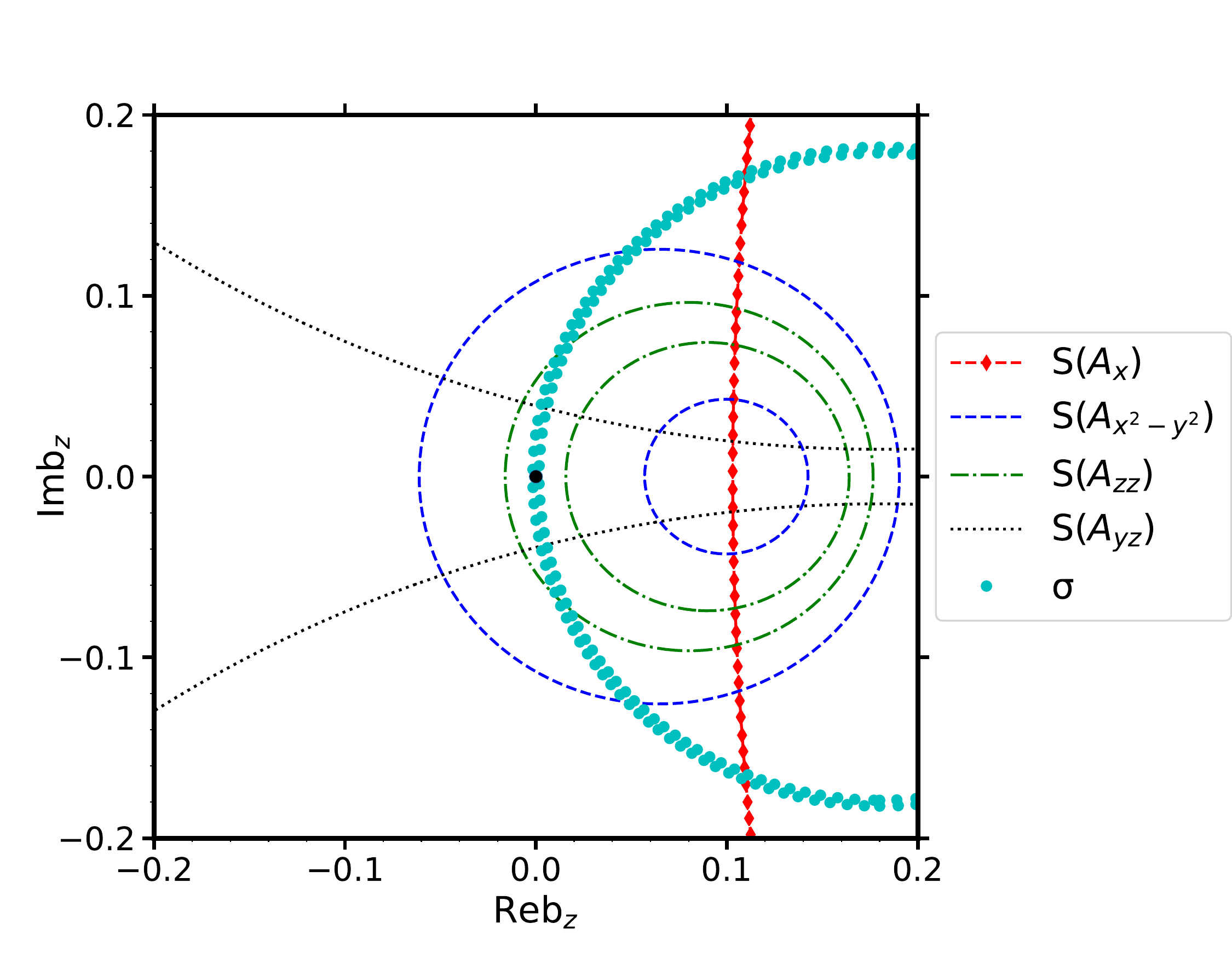}
%\caption{First subfigure} \label{fig:a}
\end{subfigure}\hspace*{\fill}
\begin{subfigure}{0.55\textwidth}
\includegraphics[width=\linewidth]{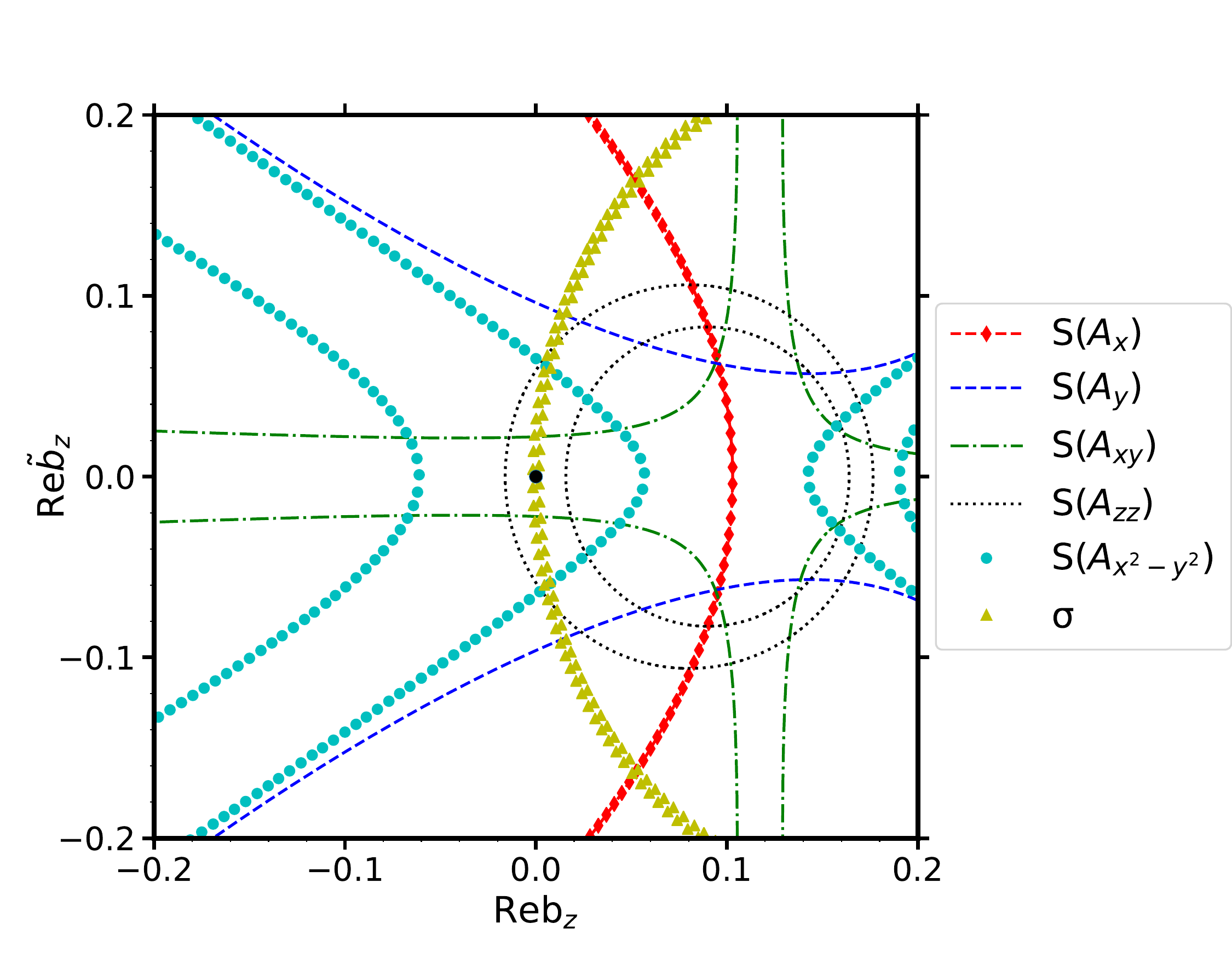}
%\caption{Second subfigure} \label{fig:b}
\end{subfigure}

\medskip
\begin{subfigure}{0.55\textwidth}
\includegraphics[width=\linewidth]{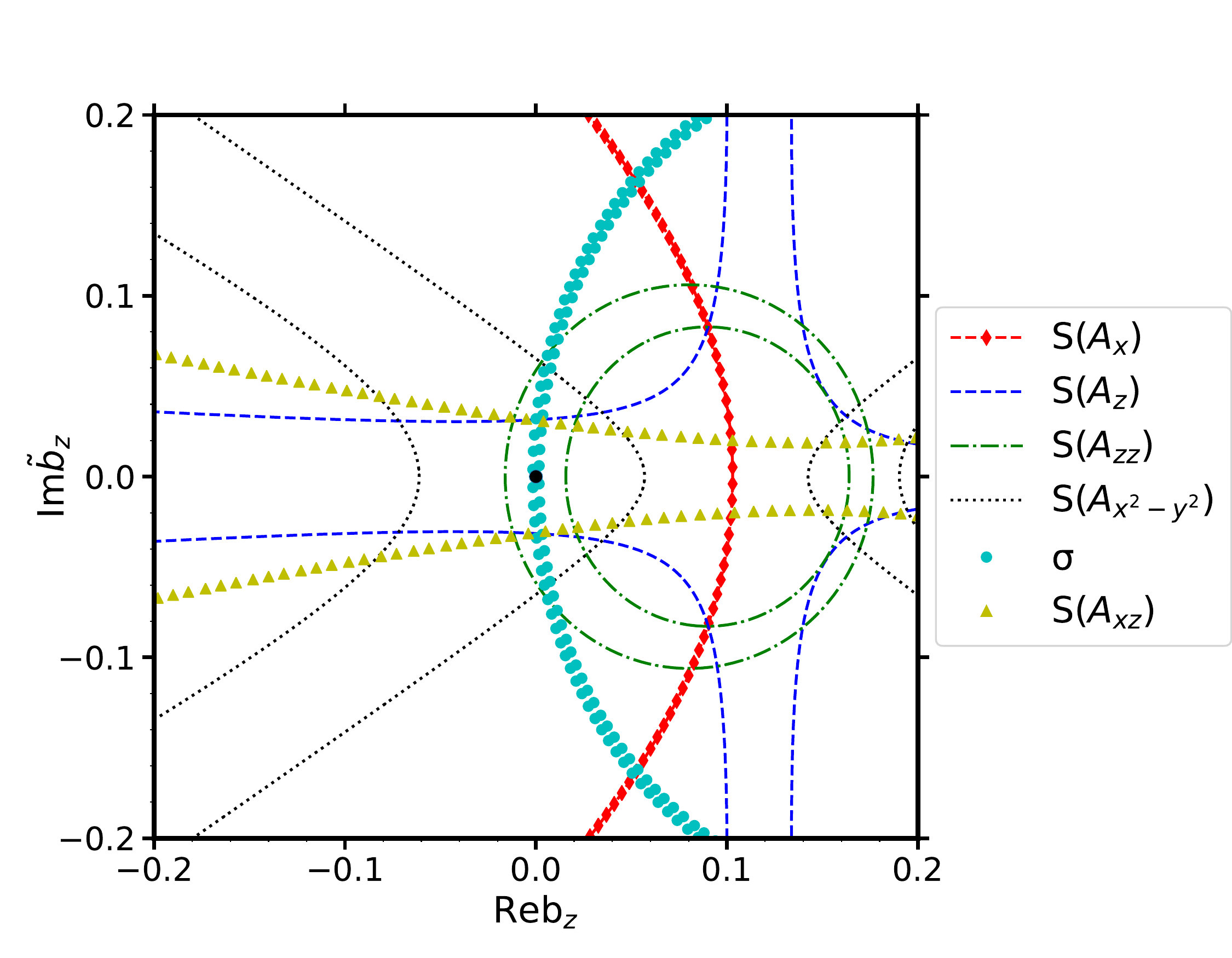}
%\caption{Third subfigure} \label{fig:c}
\end{subfigure}\hspace*{\fill}
\begin{subfigure}{0.55\textwidth}
\includegraphics[width=\linewidth]{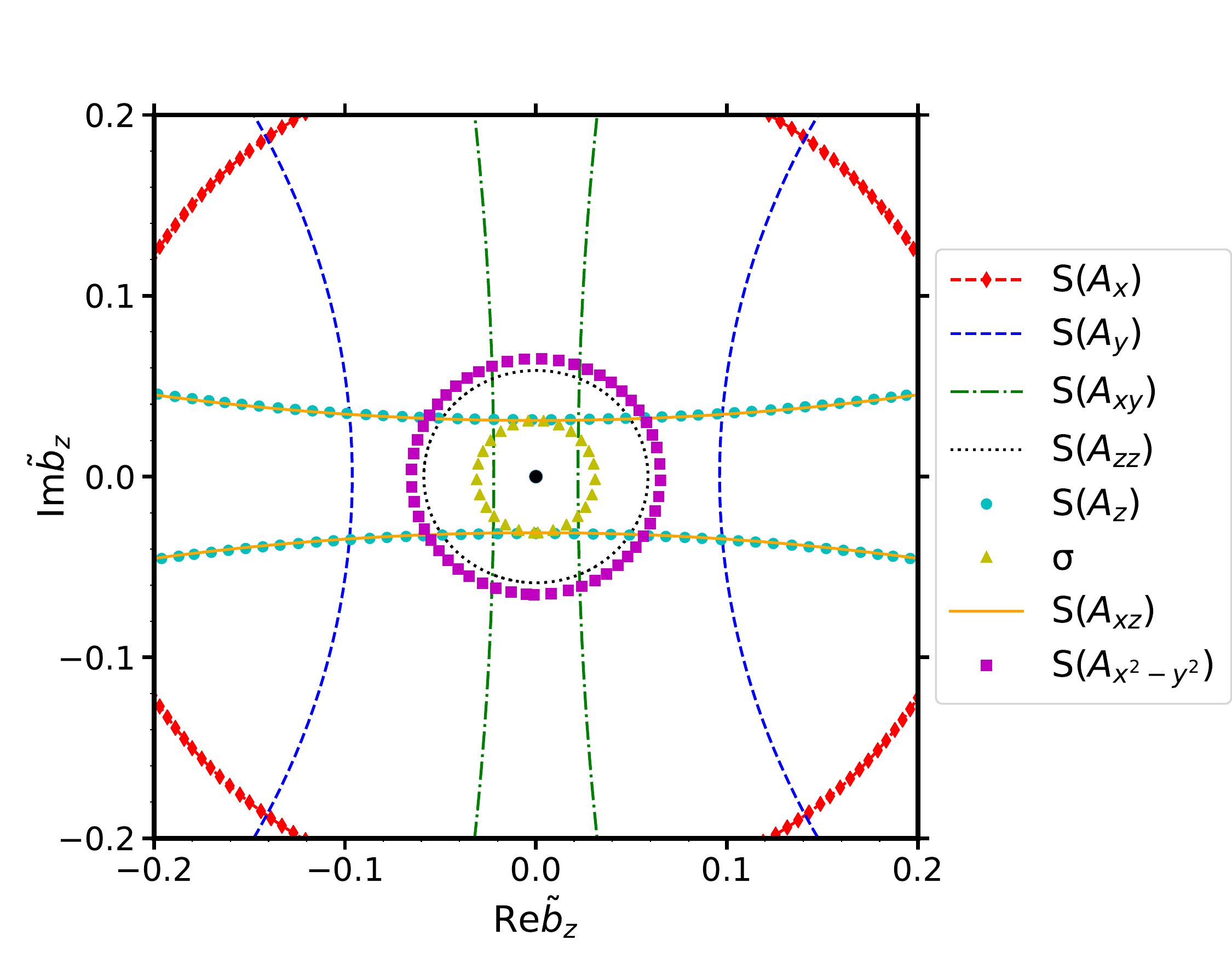}
%\caption{Fourth subfigure} \label{fig:d}
\end{subfigure}

\medskip
\begin{subfigure}{0.55\textwidth}
\includegraphics[width=\linewidth]{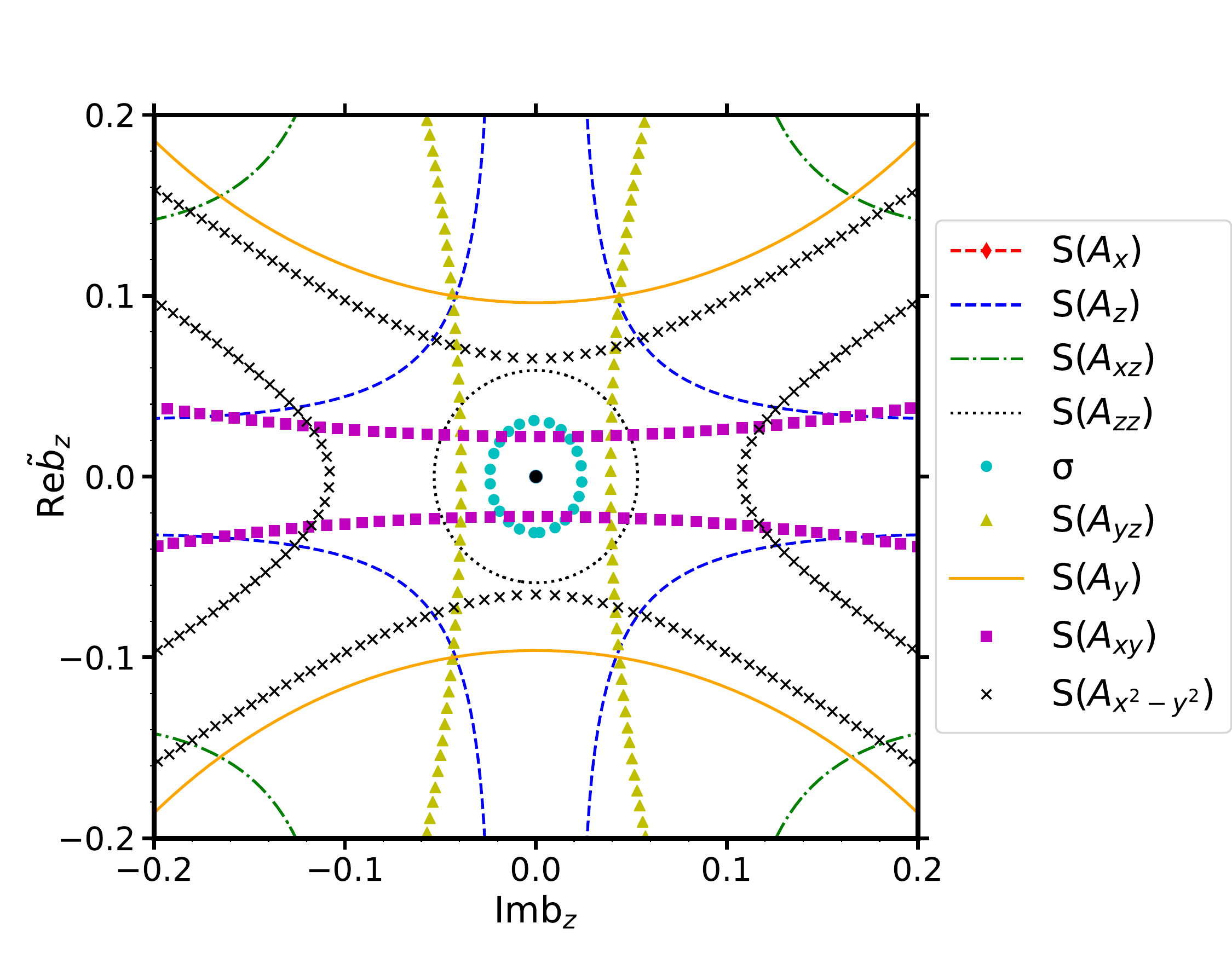}
%\caption{Fifth subfigure} \label{fig:e}
\end{subfigure}\hspace*{\fill}
\begin{subfigure}{0.55\textwidth}
\includegraphics[width=\linewidth]{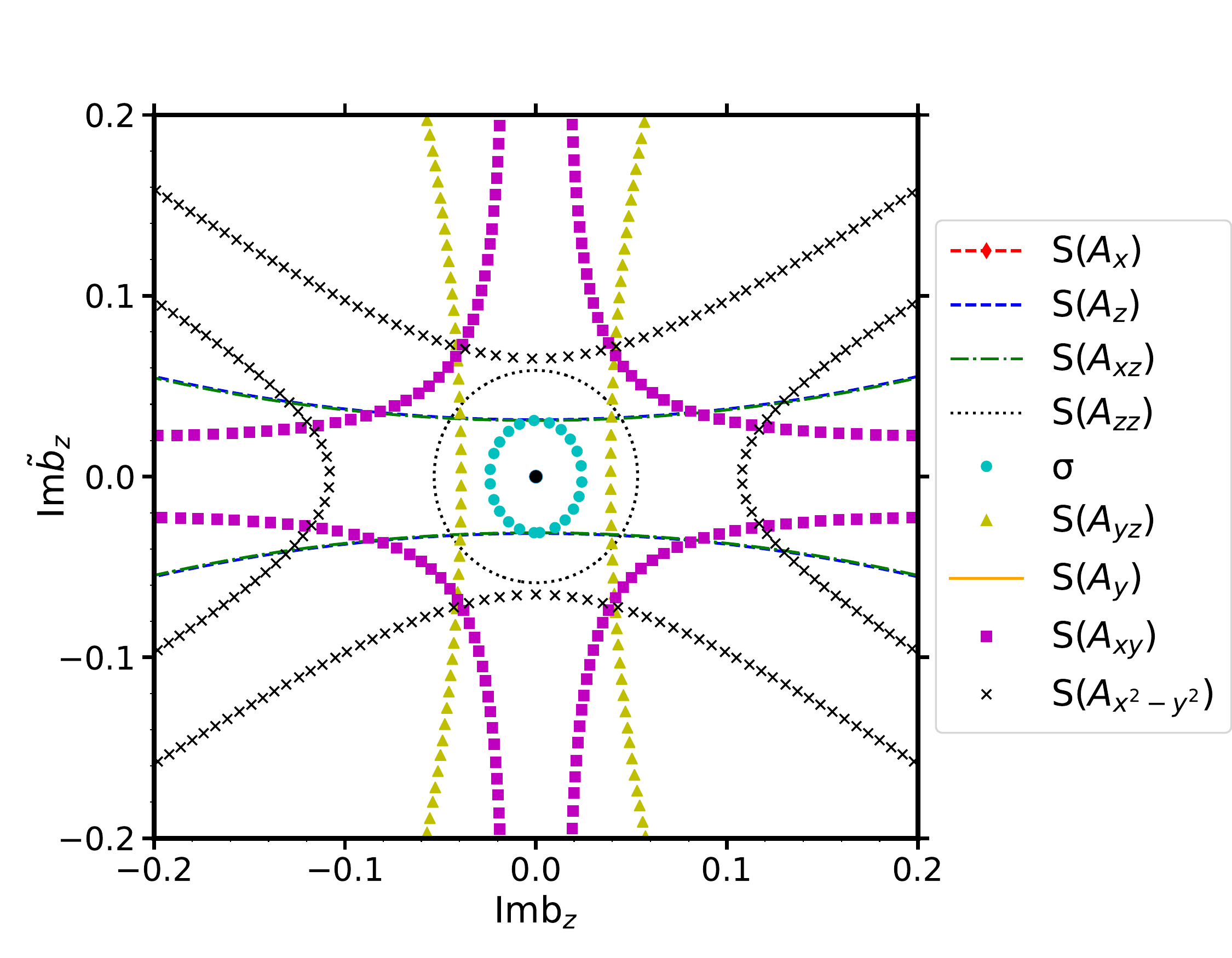}
%\caption{Sixth subfigure} \label{fig:f}
\end{subfigure}
 \caption{\label{fig:2}$1 \sigma$ sensitivity contours for cross-section and asymmetries obtained by varying two parameters simultaneously.  The black dot in the middle of the plots represents the SM value. }
%\caption{My complicated figure} \label{fig:1}
\end{figure}

So far, we obtained a limit on each coupling assuming all the other
couplings to be zero. Ideally, we would like to place a limit on each
coupling without making any assumptions on the remaining couplings.
In practice this would involve making a simultaneous fit to several observables
varying all the couplings. This is not only cumbersome, it would also
require a large data set. We therefore now consider simultaneous limits
which may be obtained by selecting a pair of couplings non-vanishing,
taking the remaining to be zero.

We vary two couplings at a time and obtain the $1 \sigma$ sensitivity \textit{i.e} $\mathcal{S}=2.3$ (or $\Delta\chi^2=2.3$) contours shown in Figure \ref{fig:2} for each observable. The black dot in the middle of the plots represents the SM value.

A first general observation in the context of deriving limits from the
contours is that the total cross section $\sigma$, which makes use of
all the events, tends to be the most sensitive observable for
measurement of all couplings. While at linear order it depends only on Re~$b_Z$,
at the quadratic order it depends on all the couplings. 
 Thus, in most cases, the best limit for all couplings is obtained from
$\sigma$.

Another observation is that as \Rebzt and \Imbzt are CP-odd couplings,
they would occur linearly in CP-odd observables. Thus, even though these
CP-odd couplings could be constrained by the cross sections or any of
the asymmetries, they would get strongest limits from the CP-odd
asymmetries $A_{y}$, $A_{xy}$, $A_{z}$ and $A_{xz}$. Of these the first
two are CPT even and would therefore constrain \Rebzt\!\!, whereas the last
two being CPT odd would constrain \Imbzt\!\!.

%Limits: 

Coming to simultaneous limits on two couplings which can be read off
 from the contour plots, the best limits on the combination \Rebz and
 \Imbz \\
 come from $\sigma$ and $A_{yz}$, and are, respectively,
$[-0.002,0.005]$ and $[-0.035,0.035]$. If, however, $A_{zz}$ is used in
place of $\sigma$, the limit on \Imbz is very similar, but the limit on
\Rebz becomes weaker. All other asymmetries give weaker limits.

Taking the sensitivity contour plots of \Rebz versus \Rebzt\!\!, the best limits
are from $\sigma$ and $A_{xy}$, the latter being linear in the
CP-odd coupling \Rebzt\!\!. These are, respectively, $[-0.002,0.003]$ and
$[-0.022,0.022]$. Again if $A_{zz}$ is used instead of $\sigma$, the
limit on \Rebz is weaker, {\it viz.}, $[-0.02,0.02]$.

In case of the contour plots of \Rebz and \Imbzt\!\!, the best limits on
\Rebz are as in the previous case, whereas the best limits on \Imbzt are $[-0.035,0.035]$, come from $\sigma$ and $A_{z}$, $A_{xz}$, the latter two being numerically very close. A similar situation holds in the case of the contour plots of \Rebzt versus \Imbzt\!\!, where the best limits on \Imbzt\!\!, {\it viz.}, $[-0.03,0.03]$ come equally from $A_{z}$ and $A_{xz}$, the best limit on \Rebzt\!\!, obtained from $A_{xy}$ is $[-0.025,0.025]$.

 The best limits on \Rebzt and \Imbz are $[-0.02,0.02]$
and $[-0.027,0.027]$ respectively, obtained from the combination of
$A_{xy}$
and $\sigma$. Lastly in the case of \Imbz versus \Imbzt contour, best limit on  \Imbz  is $[-0.025,0.025]$, comes from $\sigma$, whereas on \Imbzt\!\!, $A_{z}$ and $A_{xz}$ which contribute almost equally, provide a stringent limit of $[-0.031,0.031]$ on it. 

We see from the above a significant feature that using 
$\sigma$ as one of the observables gives a stringent limit for all the 
couplings involved. We also see that the best limits on \Rebz is of the order
of $2-5 \times 10^{-3}$ in magnitude from all relevant pairs of
observables. This may be compared to the limit $0.7 \times 10^{-3}$
obtained when only \Rebz is taken as non-zero, as seen from Tables 1 and
2. Similarly, the best
limits on \Imbz from simultaneous measurement of two observables varies
between $25 \times 10^{-3}$ and $35 \times 10^{-3}$, as compared to the
best individual limit of around $16\times 10^{-3}$. For \Rebzt the best
simultaneous limits are $20-27 \times 10^{-3}$, the best individual
limit being $9.5 \times 10^{-3}$. Likewise, the best simultaneous limits
on \Imbzt vary between $30\times 10^{-3}$ and $35 \times 10^{-3}$,
whereas the best individual limit is around $13 \times 10^{-3}$.

\section{Conclusions and discussion}\label{sec5}
The measurement of couplings of the Higgs Boson to all other SM particles is an essential test of the SM. In this work, we study the form and magnitude of the tensor structure of the couplings of the Higgs boson to a pair of $Z$ bosons at the LHC
with the help of the polarization observables of the $Z$.
We estimate sensitivities of these polarization observables by  adopting the formalism which connects angular asymmetries of charged leptons from $Z$ decay to the  polarization parameters of the $Z$. We first calculate the $Z$ polarization parameters  using the spin density matrix elements evaluated at production level and then obtain various angular asymmetries corresponding to these parameters.

We have restricted ourselves to tree-level calculations. To our
knowledge, non-leading order (NLO) contributions to the process with polarized
$Z$ have not been calculated so far. However, we expect that asymmetries
which we make use of, being ratios of cross sections, will be less
sensitive to  NLO corrections.

 We see that the $1\sigma$ limits obtained on the real parts of the couplings are of the order of a few times $10^{-3}$ and an order of magnitude higher for the imaginary parts. We show that the LHC at c.m energy
$\sqrt{s}=14\text{~TeV}$ with integrated luminosity $\int
\mathcal{L}dt=1000$ $\text{fb}^{-1}$ could provide a limit on the CP
conserving couplings $\text{Re}~b_{Z}$ in the interval  $[-0.7,0.7]\times
10^{-3}$ and $\text{Im}~b_{Z}$ in the interval  $[-15.9,15.9]\times
10^{-3}$. Similarly the CP violating couplings, $\text{Re}~\tilde{b}_{Z}$ and $\text{Im}~\tilde{b}_{Z}$ get a best bound of $\vert\text{Re}~\tilde{b}_{Z}\vert\leq9.54\times 10^{-3}$ and $\vert\text{Im}~\tilde{b}_{Z}\vert\leq13.3\times 10^{-3}$ respectively. These limits are obtained by varying one coupling at a time. With two non-zero couplings, we observe a slight weakening of bounds on all the anomalous couplings as can be expected.

We have  not considered Higgs decays, which do not affect the
polarization parameters and asymmetries of the $Z$. The effect of Higgs
decay on the sensitivities can be estimated by multiplying the SM cross
section by the Higgs branching ratio and detection efficiencies in Eqns. (\ref{limit_A}) and (\ref{limit-cs}).

Associated Higgs production with $V=W,Z$ and with $H$ decaying into $b \bar{b}$ and $V$ decaying to 0, 1 and 2 leptons has been observed by both ATLAS\cite{Aaboud:2018zhk} and CMS collaborations\cite{Sirunyan:2018kst} at close to $5\sigma$ CL. Also, as shown in \cite{Goncalves:2018fvn}, a measurement
of (SM) $Z$ polarization parameters themselves can help suppress backgrounds and enhance the
signal sensitivity in the $Z(\ell^+ \ell^-) H(b \bar{b})$ decay mode.
%For the two-charged lepton channel $Z(l^{+}l^{-})H(b \bar{b})$ considered in our work, the significance reduces to 3.4 $\sigma$ (1.9 $\sigma$) at ATLAS (CMS) at 13 TeV cm energy and 79.8 fb$^{-1}$ (41.3 fb$^{-1}$) of data respectively. However it is seen in \cite{Goncalves:2018fvn} that how information on $Z$ boson polarization-that is not yet considered in the current collider analyses, improves  the reduced sensitivity.}
A full scale analysis using an event generator coupled with all appropriate cuts and detection efficiencies relevant to the decay channels of the $Z$ and Higgs, as used in \cite{Aaboud:2018zhk, Sirunyan:2018kst, Goncalves:2018fvn} with be able to refine the actual sensitivities that we have obtained for all the eight BSM polarization asymmetries.

\vskip .2cm
\noindent {\bf Acknowledgement}: SDR acknowledges partial support from the Department of
Science and Technology, India, under the J.C. Bose National
Fellowship programme, Grant No. SR/SB/JCB-42/2009, and from the Indian National Science Academy,
New Delhi, under the Senior Scientist Programme.

%\section*{References}

\end{document}